\documentclass[byrevtex,pre,showpacs]{revtex4}
\usepackage{graphicx}
\usepackage{amsmath}
\newcommand{\fig}[1]{Fig.~\ref{#1}}
\newcommand{\eq}[1]{Eq.~(\ref{#1})}
\begin{document}
\title{Cluster diffusion at the gelation point}

\author{Sune N\o rh\o j Jespersen}
\email[]{sjespers@sfu.ca}
\affiliation{Department of physics, Simon Fraser University, Burnaby,
British Columbia, Canada V5A 1S6}

\date{\today}

\begin{abstract}
We use molecular dynamics simulations to study a model of the gelation
transition with a dynamic bond forming procedure. After establishing
evidence for $3D$ percolation as the static universality class, we
turn our attention to the dynamics of clusters at the gelation point,
focusing in particular on the behavior of the diffusion constant $D(s)$ as a function of
cluster size $s$. We find a very clear power law behavior $D(s)\sim s^{-k}$ with a
non-trivial exponent $k\approx 0.69$.
\end{abstract}

\maketitle

%%%%%%%%%%%%%%%%%%%%%%%%%%%%%%%%%%%%%%%%%%%%%%%%%%
%%%%%%%%%%%%%%%%%%%%%%%%%%%%%%%%%%%%%%%%%%%%%%%%%%
\section{Introduction}
%%%%%%%%%%%%%%%%%%%%%%%%%%%%%%%%%%%%%%%%%%%%%%%%%%
%%%%%%%%%%%%%%%%%%%%%%%%%%%%%%%%%%%%%%%%%%%%%%%%%%
Gelation is a phenomenon that continues to receive wide interest in
the physics community \cite{vernon01,broderix02,gado02}. It is a
problem with relevance both from a fundamental physics point of view,
but also from a more application oriented view due to the
many uses of gels in industry. Despite a long-standing effort the
underlying physics has still not received a complete explanation, in
particular as far as the dynamical quantities are concerned.

In a solution of polyfunctional monomers irreversible chemical bonds may be formed via some
external influence, e.g.\ irradiation with $\gamma$-rays \cite{schosseler84,schosseler85}. As the
density $p$ of bonds increases, the resulting macromolecules grow larger and
larger, until at a critical density $p_c$ (the gelation point) there
is a cluster
spanning the entire system, and a gel has formed. There are a number
of static/thermodynamic quantities characterizing this transition,
e.g.\ cluster size distribution, largest cluster, correlation length and
radius of gyration, that in many cases are believed to be well described by
the $3$ dimensional ($3D$) percolation universality class \cite{adam96}. Besides the
appearance of a wide distribution of cluster sizes, the presence of
the gel fundamentally changes many of the dynamic properties as
well. For example it has been found that the viscosity $\eta$
diverges as $(p_c-p)^{-s}$ when $p\to p_c$, but there is still no general
agreement on the value of the exponent $s$. However there are many other
quantities that probe the unique dynamics at the gelation point, and
here we consider, among other things, the diffusion of clusters of
different sizes. Diffusion in general is well known to be sensitive to
geometry and topology as well as to the interactions with the
environment, and in the case of gelation we expect the motion of the
individual macromolecules to be strongly affected by their interaction
with other macromolecules with a wide distribution of sizes.

Here we consider a model for polycondensation: using
molecular dynamics we simulate a solution of hexafunctional monomers interacting via
the Lennard-Jones potential and allow the particles to form permanent
chemical bonds if they get close enough together. After defining the
model in Sec.~\ref{sec.model}, we 
determine the static universality class by considering the behavior
of several characteristic structural quantities aided by finite size
scaling in Sec.~\ref{sec.static}. Here we also measure
the radius of gyration of the clusters as a function of cluster size
at the gelation point allowing us to determine the fractal dimension. In
Sec.~\ref{sec.dif} we focus on the behavior of the
self-diffusion constant of clusters at the gelation point, a
quantity which has hitherto received little attention (see however \cite{gould81}). Finally in
Sec.~\ref{sec.con} we give our conclusions.

%%%%%%%%%%%%%%%%%%%%%%%%%%%%%%%%%%%%%%%%%%%%%%%%%
%%%%%%%%%%%%%%%%%%%%%%%%%%%%%%%%%%%%%%%%%%%%%%%%%
\section{\label{sec.model} Model}
%%%%%%%%%%%%%%%%%%%%%%%%%%%%%%%%%%%%%%%%%%%%%%%%%
%%%%%%%%%%%%%%%%%%%%%%%%%%%%%%%%%%%%%%%%%%%%%%%%%

Our system is composed of $N=L^3$ ($L=10,15$ and $20$) particles
interacting pairwise through the shifted
Lennard-Jones potential 
\begin{equation}
U({\mathbf r})=\begin{cases}
U_{LJ}(r)-U_{LJ}(2.5\sigma), & r \leq 2.5\sigma\\
0 & \text{otherwise,}\end{cases}
\end{equation}
where
$U_{LJ}(r)=4\epsilon((\sigma/r)^{12}-(\sigma/r)^{6})$. All of our
simulations are $3D$ constant energy simulations corresponding to an
average temperature of $k_BT/\epsilon\approx 1$ and density
$\Phi=0.8\sigma^{-3}$. These 
choices ensure that the system is in the liquid-phase region of the
phase-diagram \cite{smit92}. We use periodic boundary conditions and a
time step of magnitude $dt=0.004 \sqrt{m\sigma^2/\epsilon}$. From a
typical equilibrium state of this liquid we let the particles form
permanent chemical bonds if they come closer than  
$r_c=2^{1/6}\sigma\approx 1.12$ (coinciding with the minimum of $U(r)$), and the
corresponding bond interaction is represented by a harmonic oscillator potential
$U_{\text{harm}}(r)=1/2 kr^2$: in our simulations we take
$k\sigma^2/\epsilon=120$. Note that this way of adding bonds breaks energy 
conservation; indeed we actually pump energy into the system when
adding bonds. With this bonding procedure cross linking is very fast ---
the average distance between the particles is comparable 
to $r_c$, so a large number of particles will be available for bonding at a given
instant. 
Each particle can bond to a total of $f=6$ other
particles, and the cross link density $p$ is then given in terms of the
number of bonds $n$ as $p=2n/fN$. Any number of particles, if fulfilling
the conditions above, can be cross linked per time step, but we halt
the bond formation when $p$ reaches a predetermined value.

%%%%%%%%%%%%%%%%%%%%%%%%%%%%%%%%%%%%%%%%%%%%%%%%%
%%%%%%%%%%%%%%%%%%%%%%%%%%%%%%%%%%%%%%%%%%%%%%%%%
\section{\label{sec.static}Static universality class}
%%%%%%%%%%%%%%%%%%%%%%%%%%%%%%%%%%%%%%%%%%%%%%%%%
%%%%%%%%%%%%%%%%%%%%%%%%%%%%%%%%%%%%%%%%%%%%%%%%%

\begin{figure}
\includegraphics{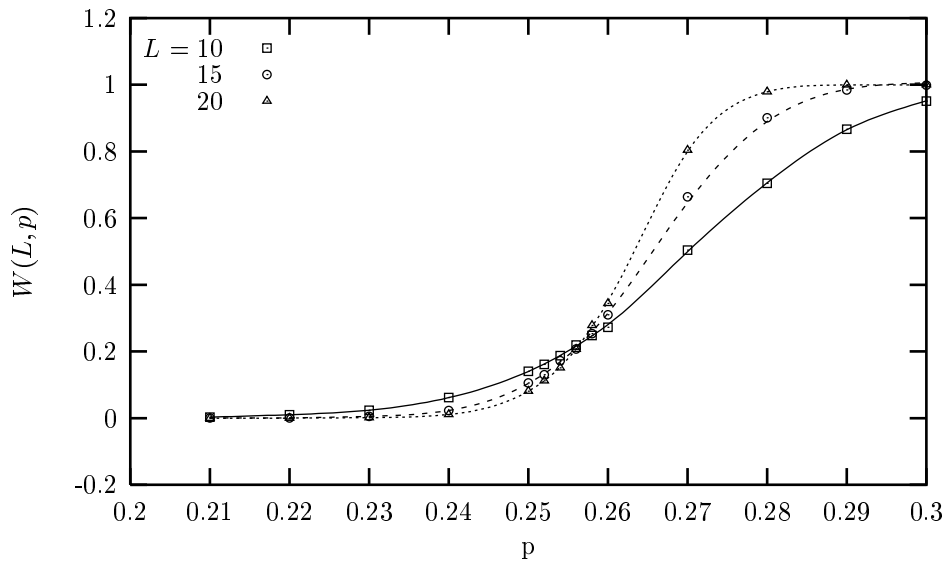}
\caption{\label{fig:pc} Fraction of systems $W(L,p)$ percolating in the
$x$-direction as a function of $p$ and for $3$ different system sizes
as indicated on the plot. The lines are guides for the eye. We
estimate $p_c=0.2565$.} 
\end{figure}

We begin by examining the geometric structure resulting from the cross
linking procedure described above. First we need to 
know the location of the geometric percolation point $p_c$, and to this end we measure
the fraction $W(L,p)$ of percolating (in one direction) systems
of size $N=L^3$ at a given cross link density $p$. In this context a
system is considered percolating when a cluster connects a particle in
the central computational box with its image in an adjacent box, and
the corresponding cluster is called a spanning cluster. In the limit
$N\to\infty$, $W(L,p)$ becomes the Heaviside step function
$\theta(p-p_c)$. However at finite $N$ for any $p>0$ below $p_c$, some
realizations will contain a spanning cluster, 
and thus $W(L,p) > 0$. Similar reasoning allows us to conclude
that $W(L,p) < 1$ for $p>p_c$. If we make the plausible assumption that $W(L,p)$
converges monotonically to $\theta(p-p_c)$ for fixed $p$, it follows that
for $L_1>L_2$, $W(L_1,p) < W(L_2,p)$ if $p<p_c$, and likewise
$W(L_1,p) > W(L_2,p)$  for any $p>p_c$. Therefore \textit{at 
and only at} $p_c$ the curves of $W(L,p)$ coincide, 
i.e.\ $W(L,p_c) = \text{const}$. In \fig{fig:pc} we plot $W(L,p)$
obtained from $5000$ realizations 
as a function of $p$ for $L=10, 15$ and $20$, and we see
that all the curves cross approximately at the same point: the
value of $p_c$ determined this way is $p_c=0.2565$.
%%%%%%%%%%%%%%%%%%%%%%%%%%%%%%%%%%%%%%%%%%%%%%%%%%
%%%%%%%%%%%%%%%%%%%%%%%%%%%%%%%%%%%%%%%%%%%%%%%%%%

\begin{figure}
\includegraphics{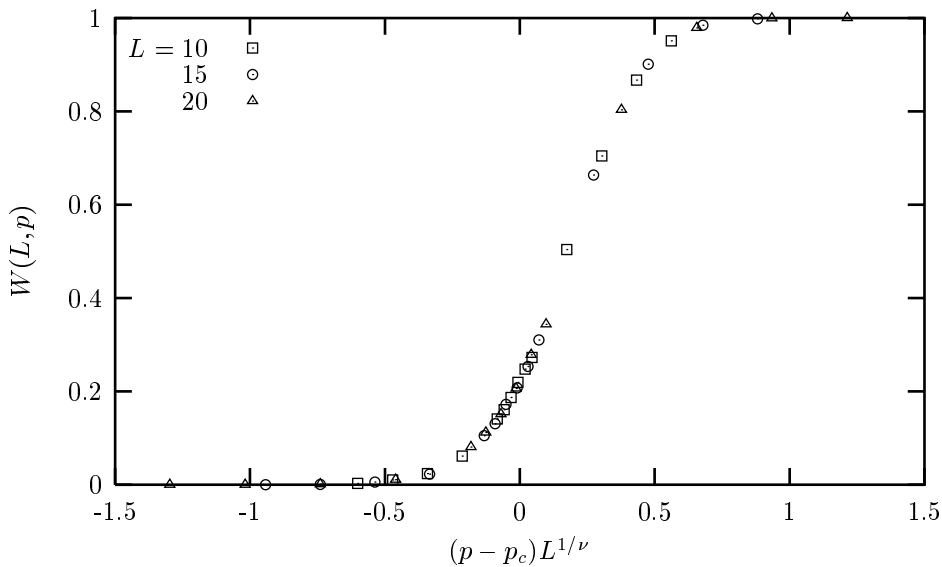}
\caption{\label{fig:scaling} Same as \fig{fig:pc}, except here
$W(L,p)$ is plotted as a function of $L^{1/\nu}(p-p_c)$ with
$p_c=0.2565$ and $\nu=0.9$. The data collapse very nicely in agreement
with finite size scaling theory.}
\end{figure}

\begin{figure}
\includegraphics{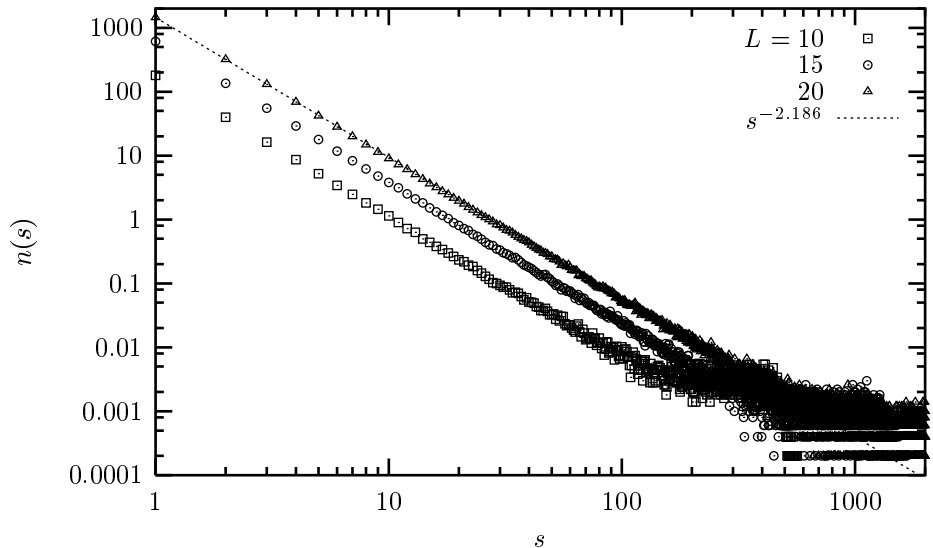}
\caption{\label{fig:clust} Here we plot the cluster size distribution
$n(s)$ as a function of $s$ for the same $3$ system sizes at $p=p_c$
and based on $5000$ samples. The straight line is a fit to a power law
to the data from the $L=20$ system.} 
\end{figure}

Finite size scaling theory predicts that $W(L,p)$ does not depend on
$L$ and $p$ separately but only on the combination $L/\xi$ (and the
sign of $p-p_c$) where
$\xi=|p-p_c|^{-\nu}$ is the correlation length 
and $\nu$ the correlation length exponent \cite{stauffer85}. Thus we
may write
\begin{equation}
\label{sizescale}
W(L,p)=f(L^{1/\nu}(p-p_c)),
\end{equation}
where $f(x)$ is a scaling function with the following limiting values:
\begin{equation}
\label{scalinglim}
\lim_{x\to\infty}f(x) = 1 \;\;\;\text{and}\;\;\; \lim_{x\to -\infty}f(x) = 0.
\end{equation}
To test this hypothesis we
replot the data for $W(L,p)$ from \fig{fig:pc} in \fig{fig:scaling} as
a function of $L^{1/\nu}(p-p_c)$ with $p_c=0.2565$ as determined above and
$\nu=0.9$.  
%%%%%%%%%%%%%%%%%%%%%%%%%%%%%%%%%%%%%%%%%%%%%%%%%%
%%%%%%%%%%%%%%%%%%%%%%%%%%%%%%%%%%%%%%%%%%%%%%%%%%
The curves collapse quite convincingly onto the same master curve with
the limiting values given in \eq{scalinglim}, and
the results are thus consistent with finite scaling theory. Moreover,
the value of $\nu$ is in very good agreement with the correlation
length exponent $0.88$ of $3D$ percolation
theory \cite{stauffer85,adam96}. Thus we believe 
that in spite of using an off-lattice dynamic bonding method from an equilibrium
liquid state, this systems belongs
to the universality class of $3D$ percolation. To further
substantiate that claim, we have also considered
the cluster size distribution at the critical point $p=p_c$
in \fig{fig:clust}. According to
standard percolation theory the number $n(s)$ of clusters of size $s$
is a power law $n(s)\sim s^{-\tau}$ with an exponent $\tau\approx
2.18$ \cite{Sahimi94}. 
%%%%%%%%%%%%%%%%%%%%%%%%%%%%%%%%%%%%%%%%%%%%%%%%%%
%%%%%%%%%%%%%%%%%%%%%%%%%%%%%%%%%%%%%%%%%%%%%%%%%%
Fitting the data from the largest system size to a power law we find
the exponent $\tau\approx 2.19$, which is within $0.5\%$ of the $3D$
percolation value quoted above. 

For the universality class of $3D$ percolation there are in principle
only two independent exponents, but to establish even more confidence
in our conclusions and because we need part of the data later, we determine
two more exponents. 

\begin{figure}
\includegraphics{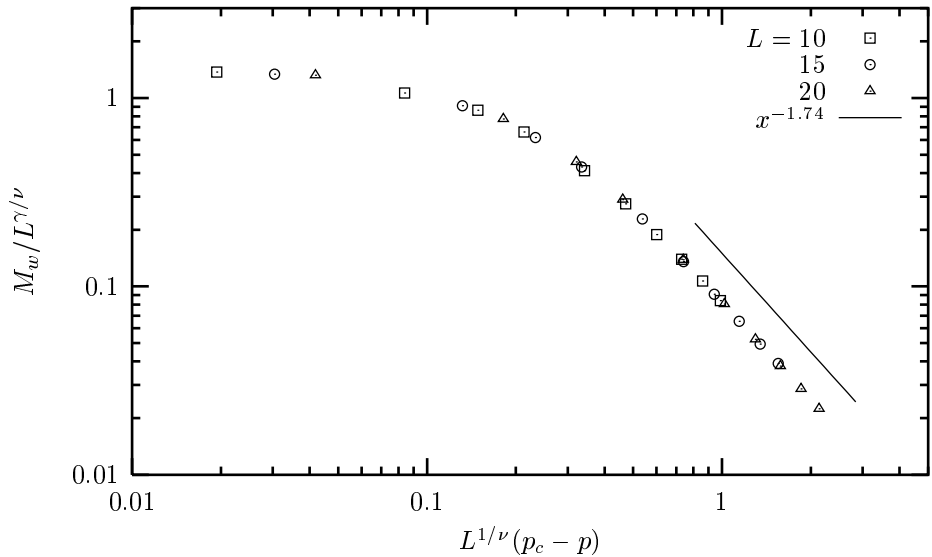}
\caption{\label{fig:mw} Scaling plot of the weight average molecular
weight $M_w$. The quality of the data collapse confirms $\gamma=1.74$
in accordance with the $3D$ percolation value.}
\end{figure}

The weight average molecular mass $M_w$ for $p<p_c$ defines the exponent
$\gamma$ by the relation $M_w\sim (p_c-p)^{-\gamma}$. Again,
for finite systems, finite size scaling theory predicts the scaling form
\begin{equation}
\label{mw}
M_w = L^{\gamma/\nu} g(L^{1/\nu}(p_c-p)),
\end{equation}
where $g(x)$ is a scaling function with the following asymptotic
behavior:
\begin{equation}
\label{gasympt}
g(x)\sim
\begin{cases}
x^{-\gamma} & x\to\infty \\
\text{const.} & x\to 0.
\end{cases}
\end{equation}
Therefore we compute $M_w$ as a
function of $p$ for different system sizes, and in \fig{fig:mw} we plot
the results in the form $M_w/L^{\gamma/\nu}$ versus
$L^{1/\nu}(p_c-p)$ with $\gamma=1.74$ being the expected $3D$
percolation value and $\nu$ and $p_c$ as determined previously. Again
there is a very nice data collapse reinforcing our statement about the
validity of $3D$ percolation exponents. 
%%%%%%%%%%%%%%%%%%%%%%%%%%%%%%%%%%%%%%%%%%%%%%%%%%
%%%%%%%%%%%%%%%%%%%%%%%%%%%%%%%%%%%%%%%%%%%%%%%%%%

\begin{figure}
\includegraphics{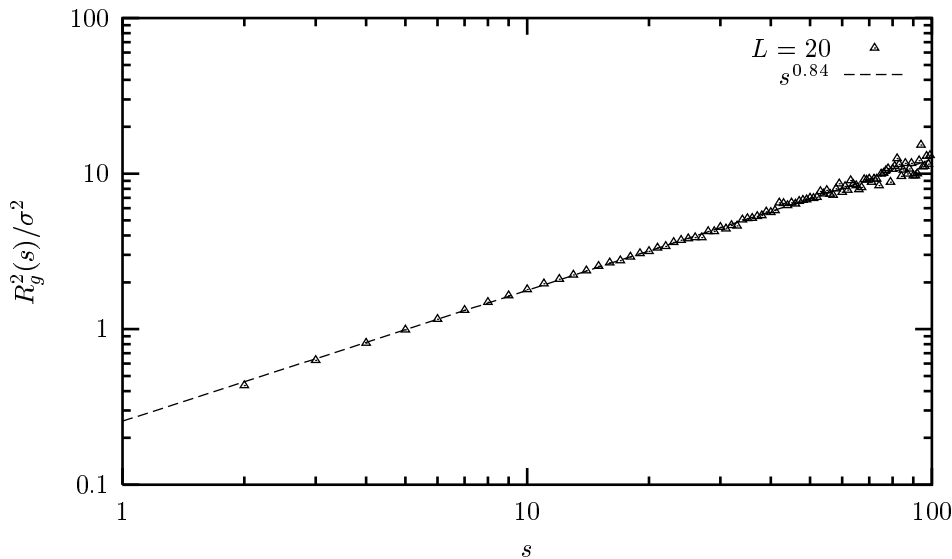}
\caption{\label{fig:rg} Radius of gyrations $R_g$ (squared) as a function of cluster
size $s$ for $L=20$ and $p=p_c$.}
\end{figure}

The last static exponent we wish to consider is the fractal dimension $D$
of the clusters at the gelation point. During the simulations we
maintain a list of all particles belonging to a 
given cluster, and this enables us to calculate the radius of gyration
of a given cluster. At a given instant $R_g(s)$ ($s\geq 2$) is found
from \cite{Plischke94}
\begin{equation}
\label{rg}
R_g^2(s)=\frac{1}{s(s-1)}\sum_{i,j=1}^s (\mathbf{r}_i-\mathbf{r}_j)^2,
\end{equation}
where the sum is over all particles in the cluster and $\mathbf{r}_i$ is
the position of particle $i$ at that instant. To calculate its thermal
average we average over time. However we have
also averaged over $100$ samples to get better statistics for the
individual cluster sizes.  We find that the
results are independent of the system size (up to a cut-off) and do
not significantly 
depend on the value of $p$ close to $p_c$, and therefore we plot here
only the results from the largest system, i.e.\ $N=20^3$. We focus on
the properties of the system at the percolation point $p=p_c$ and the
corresponding data are plotted in \fig{fig:rg}.
%%%%%%%%%%%%%%%%%%%%%%%%%%%%%%%%%%%%%%%%%%%%%%%%%%
%%%%%%%%%%%%%%%%%%%%%%%%%%%%%%%%%%%%%%%%%%%%%%%%%%

There is a clear power law behavior $R_g^2(s)\sim s^{2/D}$, and from a
least squares fit we deduce the value $D=2.38$. Here $D$ is the fractal
dimension of the clusters (since $s\sim R_g^D$) at the gelation point,
and for this exponent $3D$ percolation theory predicts the
value $D\approx 2.5$. Our value is about $5\%$ lower, but this small
discrepancy may be due to lack of data for sufficiently large $s$. However it
may also be due to the interaction of the polymers with one another
and with the left over monomers.

Therefore regarding the static exponents we presented evidence that our
model of the gelation transition, which uses a Lennard-Jones force-shifted
intermolecular potential and a liquid state bond forming procedure, is
consistent with the universality class of $3D$ percolation.

%%%%%%%%%%%%%%%%%%%%%%%%%%%%%%%%%%%%%%%%%%%%%%%%%
%%%%%%%%%%%%%%%%%%%%%%%%%%%%%%%%%%%%%%%%%%%%%%%%%
\section{\label{sec.dif}Cluster diffusion}
%%%%%%%%%%%%%%%%%%%%%%%%%%%%%%%%%%%%%%%%%%%%%%%%%
%%%%%%%%%%%%%%%%%%%%%%%%%%%%%%%%%%%%%%%%%%%%%%%%%

We now turn to the measurement of a dynamic cluster specific quantity,
namely the diffusion coefficient $D(s)$ as a function of cluster size $s$ at the
percolation threshold $p=p_c$. We measure the mean square fluctuations of the center
of mass position $R^{\text{CM}}_s(t)$ for each cluster of size
$s$, and for large times we find a linear behavior
\begin{equation}
\label{diffusion}
\langle (R^{\text{CM}}_s(t)-R^{\text{CM}}_s(0))^2\rangle \sim 6 D(s) t,
\end{equation}
allowing us to extract the diffusion constant $D(s)$ as a function of cluster mass
$s$. Again the average $\langle\cdots\rangle$ is performed over both
time (thermal average) and over realizations, and here we have found
it sufficient to use $80$ samples. 

\begin{figure}
\includegraphics[width=\textwidth]{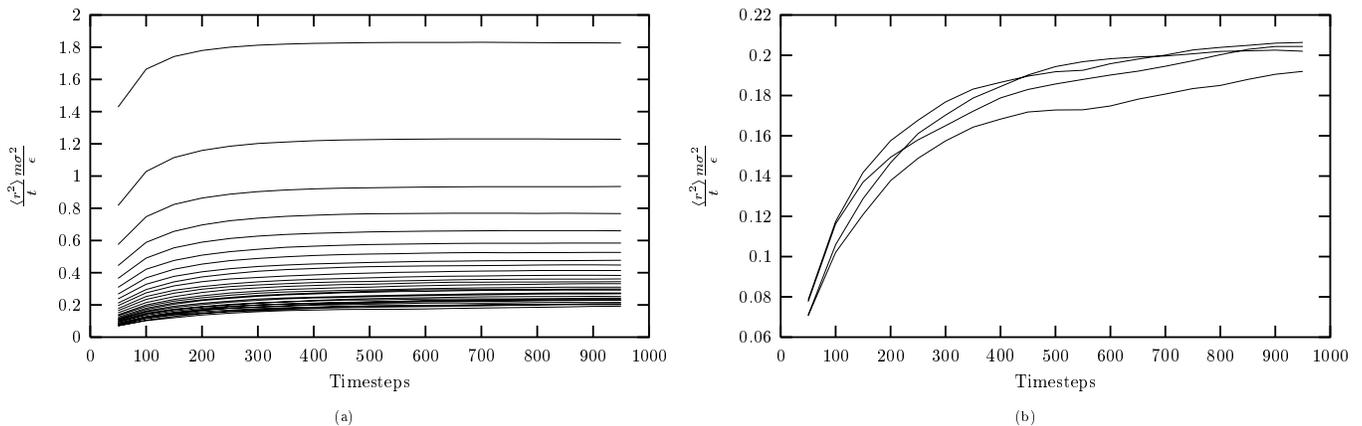}
\caption{\label{fig:rel} Mean square displacements divided by time
for the individual clusters of the $N=20^3$ system at $p=p_c$. (a) Clusters of size
$1\ldots 29$ from top to bottom. The asymptotic convergence to
horizontal lines indicates normal diffusion, c.f.\
\eq{diffusion}. (b) Close up of (a) for the $4$ largest clusters. They
have all reached an almost constant value, except perhaps the largest
cluster, which therefore represents the ``worst case''.}
\end{figure}

To demonstrate the asymptotic normal diffusion we plot for the
$N=20^3$ system in \fig{fig:rel}.a
$\langle (R^{\text{CM}}_s(t)-R^{\text{CM}}_s(0))^2\rangle/t$ as a function
of time steps for several cluster sizes $s=1\ldots 29$, and part (b)
shows a magnification for the largest clusters $s=26\ldots 29$. As can be seen
all the curves approach constant values for large times, and it is
from these values that we calculate the diffusion constant using
\eq{diffusion}. We have checked that the results below do not change
if we extend the length of the sampling time to much longer times
($5000$ timesteps).
%%%%%%%%%%%%%%%%%%%%%%%%%%%%%%%%%%%%%%%%%%%%%%%%%%
%%%%%%%%%%%%%%%%%%%%%%%%%%%%%%%%%%%%%%%%%%%%%%%%%%

\begin{figure}
\includegraphics{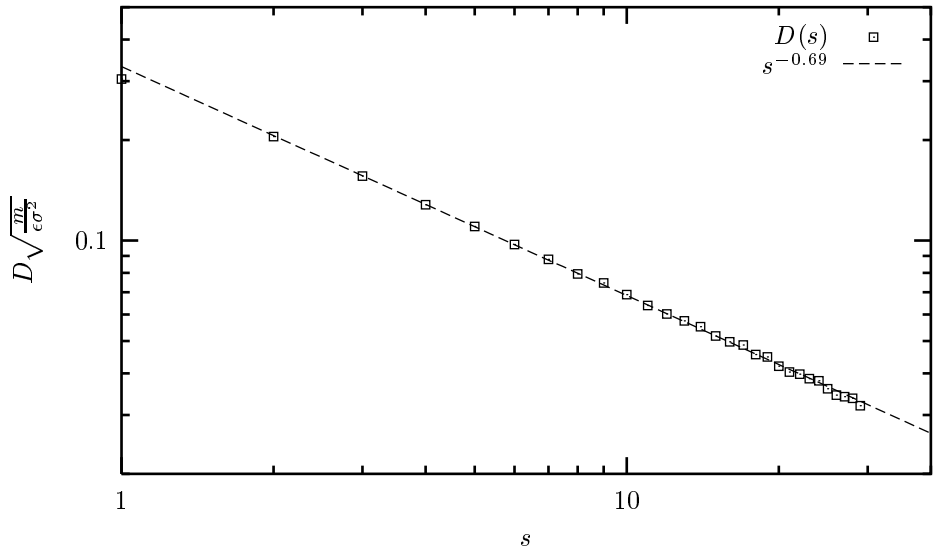}
\caption{\label{fig:Dvn} Diffusion constant $D(s)$ as a function of
the mass $s$ (or size) of the cluster for the $N=20^3$ system at $p=p_c$. The solid line is
a power law $s^{-0.69}$ obtained from a least squares fit to the data
in the region $\lbrack 3\colon\! 30 \rbrack$.}
\end{figure}

For monodisperse polymers in a dilute solution, Rouse dynamics
predicts the power law dependence $D(s)\sim s^{-1}$. Taking
hydrodynamic interactions into account, Zimm dynamics predicts
$D(s)\sim R_g^{-1}\sim s^{-1/D}=s^{-0.4}$ \cite{Doi86}. For the
present model we observe a very clear power law $D(s)\sim s^{-k}$ as can be seen in
\fig{fig:Dvn}, where we have plotted $D(s)$ as a function of $s$. We
consider again only the largest system size $N=20^3$ at $p=p_c$, and $D(s)$
is found from \eq{diffusion} based on the data from \fig{fig:rel}. The
exponent of the power law is approximately $k\approx 0.69$, and is
thus described neither by pure Rouse nor Zimm dynamics.
%%%%%%%%%%%%%%%%%%%%%%%%%%%%%%%%%%%%%%%%%%%%%%%%%%
%%%%%%%%%%%%%%%%%%%%%%%%%%%%%%%%%%%%%%%%%%%%%%%%%%
Instead the exponent is intermediate between the values for Rouse and
Zimm dynamics.

\begin{figure}
\includegraphics{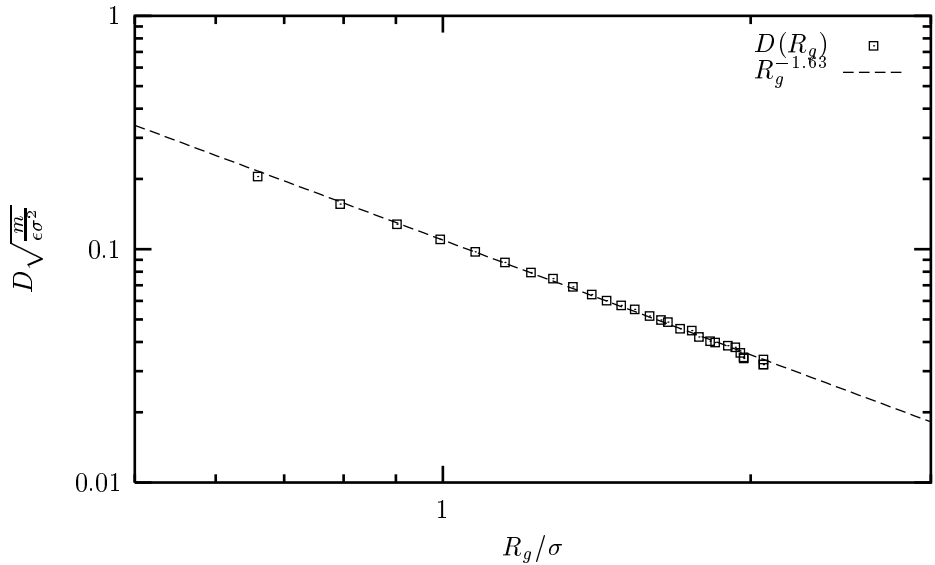}
\caption{\label{fig:Dvrg} Diffusion constant as a function of radius of
gyration for $N=20^3$ and $p=p_c$.}
\end{figure}

Having measured the radius of gyration in \fig{fig:rg}, we can convert
the result above into a dependence of $D$ on $R_g$: $D(R_g)\sim
s^{-0.69}\sim R_g^{-0.69D}=R_g^{1.64}$, and the 
corresponding data are plotted in \fig{fig:Dvrg}. Also shown is a power
law fit to the data yielding an exponent value of $1.63$. Again we
emphasize the clear deviation from a Stokes' type law, $D(R_g)\sim
R_g^{-1}$, as obtained from Zimm dynamics for polymers. Moreover our
result is in sharp contrast to a corresponding result in a 
recent paper \cite{gado00} for a lattice model of gelation in which
it was found that $D(R_g)\sim R_g^{-2.4}$ using bond fluctuation dynamics. 
%%%%%%%%%%%%%%%%%%%%%%%%%%%%%%%%%%%%%%%%%%%%%%%%%%
%%%%%%%%%%%%%%%%%%%%%%%%%%%%%%%%%%%%%%%%%%%%%%%%%%

%%%%%%%%%%%%%%%%%%%%%%%%%%%%%%%%%%%%%%%%%%%%%%%%%%
%%%%%%%%%%%%%%%%%%%%%%%%%%%%%%%%%%%%%%%%%%%%%%%%%%
\section{\label{sec.con}Conclusions}
%%%%%%%%%%%%%%%%%%%%%%%%%%%%%%%%%%%%%%%%%%%%%%%%%%
%%%%%%%%%%%%%%%%%%%%%%%%%%%%%%%%%%%%%%%%%%%%%%%%%%
Using extensive molecular dynamics simulations of a system of Lennard-Jones
particles we studied a specific model for the gelation
transition. An important feature of our model is that particles are
allowed to form permanent chemical bonds from the Lennard-Jones liquid state when they get
close together. For this model we determined a number of exponents
characterizing the geometry and polydispersity of the system, and this
information allowed us to conclude that $3D$ percolation theory is
the appropriate static universality class. We went on to discuss the radius
of gyration of the clusters at the gel point resulting in the
determination of their fractal dimension. Afterwards we
focused on the self-diffusion constant of individual clusters as a 
function of their size. We found strong indications for a power law
behavior with an exponent described neither by Rouse nor
Zimm-dynamics. In future work we intend to study e.g.\ the shear
viscosity of this model in order to measure the corresponding dynamic
exponents. As mentioned in the introduction there is
considerable disagreement as to the values of these exponents, and
numerical results on models such as the present one, will allow us to
discriminate between some of the proposed dynamic universality classes
for gelation.

%%%%%%%%%%%%%%%%%%%%%%%%%%%%%%%%%%%%%%%%%%%%%%%%%%
\begin{acknowledgments}
The author wishes to thank M. Plischke and D. Vernon for helpful
discussions. Financial support from the Danish National Research
Council grant $21$-$01$-$0335$ and from NSERC of Canada is gratefully
acknowledged.  
\end{acknowledgments}

\bibliography{ms0.bbl}

\end{document}